# Anomalous thermal expansion in one-dimensional transition metal cyanides: Behavior of the trimetallic cyanide $Cu_{1/3}Ag_{1/3}Au_{1/3}CN$

Stella d'Ambrumenil[1,2], Mohamed Zbiri[1*], Simon J. Hibble[3], Ann M. Chippindale[2], Dean S. Keeble[4], Camille Wright[2], Nicholas H. Rees[5]

[1]Institut Laue-Langevin, 71 avenue des Martyrs, Grenoble Cedex 9, 38042, France.
[2]Department of Chemistry, University of Reading, Whiteknights, Reading, RG6 6AD, United Kingdom.
[3]Chemistry Teaching Laboratory, Department of Chemistry, University of Oxford, South Parks Road, Oxford, OX1 3PS, United Kingdom.
[4]Diamond Light Source, Harwell Campus, Oxfordshire, OX11 0DE, United Kingdom.
[5]Department of Chemistry, University of Oxford, Mansfield Road, Oxford OX1 3TA, United Kingdom.

*zbiri@ill.fr

## Abstract

The structural dynamics of a quasi-one-dimensional (1D) mixed-metal cyanide, $Cu_{1/3}Ag_{1/3}Au_{1/3}CN$, with intriguing thermal properties is explored. All the current known related compounds with straight-chain structures, such as group 11 cyanides CuCN, AgCN, AuCN and bimetallic cyanides $M_xM'_{1-x}CN$ (M, M' = Cu, Ag, Au), exhibit 1D negative thermal expansion (NTE) along the chains and positive thermal expansion (PTE) perpendicular to them. $Cu_{1/3}Ag_{1/3}Au_{1/3}CN$ exhibits similar PTE perpendicular to the chains, however PTE, rather than NTE, is also observed along the chains. In order to understand the origin of this unexpected behavior, inelastic neutron scattering (INS) measurements were carried out, underpinned by lattice-dynamical density-functional-theory (DFT) calculations. Synchrotron-based pair-distribution-function (PDF) analysis and $^{13}C$ solid-state nuclear-magnetic-resonance (SSNMR) measurements were also performed to build an input structural model for the lattice dynamical study. The results indicate that transverse motions of the metal ions are responsible for the PTE perpendicular to the chains, as is the case for the related group 11 cyanides. However NTE along the chain due to the *tension effect* of these transverse motions is not observed. As there are different metal-to-cyanide bond lengths in $Cu_{1/3}Ag_{1/3}Au_{1/3}CN$, the metals in neighboring chains cannot all be truly co-planar in a straight-chain model. For this system, DFT-based phonon calculations predict small PTE along the chain due to low-energy chain-slipping modes induced by a *bond-rotation effect* on the weak metallophilic bonds. However the observed PTE is greater than that predicted with the straight-chain model. Small bends in the chain provide an alternative explanation for thermal behavior. These would mitigate the *tension effect* induced by the transverse motions of the metals and, as temperature increases and the chains move further apart, a straightening could occur resulting in the observed PTE. This hypothesis is further supported by unusual evolution in the phonon spectra, which suggest small changes in local symmetry with temperature.

# I. INTRODUCTION

Negative thermal expansion (NTE) is a counterintuitive phenomenon which can manifest in 1, 2, or 3 dimensions [1]. Understanding the dynamical mechanisms behind the phenomenon provides valuable information that helps design other NTE compounds or even zero-thermal-expansion composites. NTE is often observed in transition-metal cyanides as the diatomic, linear, cyanide ligand provides a degree of flexibility that allows for low-energy transverse modes that give rise to the *tension effect* [1]. Well-known examples include high-temperature (HT) CuCN, $Ni(CN)_2$ and $Zn(CN)_2$ [2–4]. In each case, partial substitution of the metal ions resulting in $Cu_xAg_{1-x}CN$, $Cu_xNi_{1-x}(CN)_4$ and $Zn_xCd_{1-x}(CN)_2$, alters the magnitude of the observed NTE [5–7].

The group 11 metal cyanides form quasi-1D materials consisting of linear M-CN chains packed together in a hexagonal arrangement [8]. AgCN, AuCN and HT-CuCN all exhibit 1D NTE parallel to their linear chains, and large positive thermal expansion (PTE) perpendicular to them [2]. The same directional behavior is observed for the bimetallic cyanides, such as $Cu_{1/2}Au_{1/2}CN$, $Ag_{1/2}Au_{1/2}CN$ and $Cu_xAg_{1-x}CN$ ($0 \leq x \leq 1$), with varying magnitudes [5]. This work explores the thermally-induced anomalous dynamical behavior of the trimetallic cyanide, $Cu_{1/3}Ag_{1/3}Au_{1/3}CN$.

The compounds mentioned above adopt one of the two structure types shown in Fig. 1 [8]. In the AuCN structure, the metal ions in neighboring chains are aligned, resulting in coplanar metal sheets. In the AgCN structure, the chains are offset by 1/3 of the interchain metal-to-metal distance. The difference between the structures of AuCN and AgCN specifically was previously attributed to the aurophilic bonding present in AuCN [8]. Structural analysis of the bimetallic cyanides showed that all the Au-containing cyanides adopt the AuCN structure [5]. In these compounds, metals have both like and unlike nearest neighbors perpendicular to the chains and there is no long-range metallic ordering within the metal planes. Nonetheless, it appears that metallophilic interactions, either between very small fractions of neighboring Au ions or between Au ions and other metals, are enough to favor the AuCN structure.

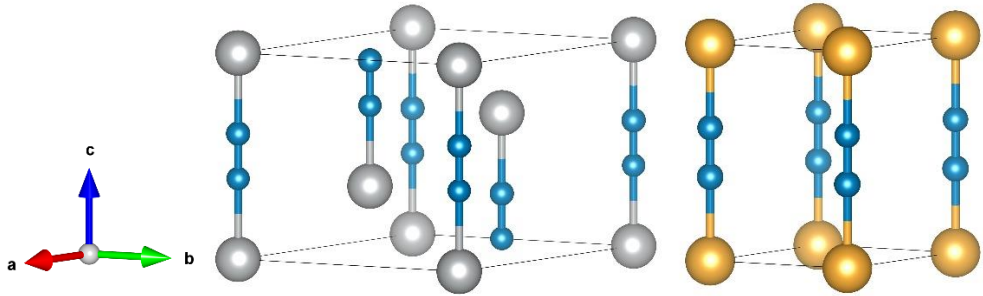

FIG. 1. Schematic illustration of the structures of AgCN (left) and AuCN (right). Due to the head-to-tail disorder of the cyanide ligand, both C and N are shown in the same color, blue.

This work focuses on the study of the underlying lattice dynamical mechanisms at the origin of the intriguing thermal expansion behavior exhibited by $Cu_{1/3}Ag_{1/3}Au_{1/3}CN$. Inelastic neutron scattering (INS) measurements, underpinned by density functional theory (DFT) calculations, were carried out to probe phonon dynamics and related properties. Synchrotron-based pair-distribution-function (PDF) analysis and $^{13}C$ solid-state nuclear-magnetic-resonance (SSNMR) measurements were performed to build an adequate input structural model for the lattice-dynamical analysis and interpretation. This combined approach provides insight into the atomistic mechanisms of thermal expansion and has previously been used to study many other transition-metal cyanides with unusual thermal properties [9–14].

## II. EXPERIMENTAL DETAILS

*Caution! Cyanide materials are toxic and must be handled with care. The addition of acid to soluble cyanides liberates highly toxic gaseous HCN. Both gaseous HCN and the aqueous washings, which contain HCN, were destroyed using alkaline hypochlorite.*

Synthesis of $Cu_{1/3}Ag_{1/3}Au_{1/3}CN$ - 2 mmol of CuCN, AgCN and AuCN, and 10 mmol of KCN were added to 8 mL of distilled water and left to stir for 30 mins to ensure the cyanides were fully dissolved. 11 mL of 1 M HNO3 was then quickly added whilst stirring vigorously. A cream precipitate formed instantly. 20 mL of distilled water was added and then decanted off. This was repeated several times. The precipitate was filtered by suction filtration and then left to air dry for 15 mins before being placed in a desiccator connected to a high vacuum line for 4 h. All reagents were used as obtained from commercial sources.

Infra-red (IR) and Raman - IR spectra were collected with a Perkin Elmer Spectrum 100 spectrometer and Raman spectra with a Renishaw InVia Raman microscope ($\lambda_{exc}$ = 785 nm).

Solid-state nuclear magnetic resonance (SSNMR) - $^{13}$C SSNMR were acquired at 100.6 MHz in a 9.4 T magnet using a DEPTH [15] sequence to suppress the probe background. The sample was packed in a 4 mm O.D. rotor and spun at 10 kHz. 1868 scans were acquired with a pulse delay of 180 seconds and an acquisition time of 12.5 ms (2000 points with a spectral width of 810 ppm). The spectrum was referenced to adamantane (the upfield methine resonance was taken to be at $\delta$ = 29.5 ppm [16] on a scale where $\delta$(TMS) = 0) as a secondary reference.

Total x-ray scattering - Total-scattering measurements were carried out on the I15-1 beamline at the Diamond Light Source, UK. The sample was loaded into a 1 mm borosilicate capillary and mounted on a spinner, where its temperature was controlled using an Oxford Instruments cryostream. The x-ray beam had a wavelength of $\lambda$ = 0.161669 Å and an energy of 76.69 keV. Calibration of the Perkin Elmer XRD 4343 CT detector was carried out using a capillary of NIST 640b silicon powder, and an empty capillary was measured for background subtraction. The integration, along with flat-field, detector-transmission, polarization and geometrical corrections were carried out using Dawn [17,18]. Lattice parameters were obtained by Pawley refinement using TOPAS [19]. The total-scattering structure function and corresponding pair distribution function (PDF) were extracted using GudrunX [20].

Inelastic neutron scattering (INS) - measurements were performed on a ~ 2-g finely-ground sample of $Cu_{1/3}Ag_{1/3}Au_{1/3}CN$, using the cold-neutron, time-of-flight, time-focusing, IN6 spectrometer (Institut Laue-Langevin, Grenoble), operating in the high-resolution mode, and offering a good signal-to-noise ratio. The IN6 spectrometer supplies a typical flux of $10^6$ n cm$^{-2}$ s$^{-1}$ on the sample, with a beam-size cross section of 3×5 cm$^2$ at the sample position. The sample was placed inside a thin-walled aluminium container and fixed to the tip of the sample stick of an orange cryofurnace. An optimized small sample thickness of 0.2 mm was used, to minimize effects such as absorption and multiple scattering. An incident wavelength of 4.14 Å was used, offering an elastic energy resolution of 0.17 meV, as determined from a standard vanadium sample. The vanadium sample was also used to calibrate the detectors and to normalise the spectra. Data were collected up to 100 meV in the up-scattering, neutron energy-gain mode, at 200, 300, 400 and 500 K. On IN6, under these conditions, the resolution function broadens with increasing neutron energy, and it can therefore be expressed as a percentage of the energy transfer. The ILL program LAMP [21] was used to carry out data reduction and treatment, including detector efficiency calibration and background subtraction. Background reduction included measuring an identical empty container in the same conditions as sample measurements. At the used shortest-available neutron wavelength on IN6, $\lambda = 4.14$ Å, the IN6 angular coverage ($\sim 10 - 114°$) corresponds to a maximum momentum transfer of $Q \sim 2.6$ Å$^{-1}$ at the elastic line.

In the incoherent approximation [22], the $Q$-averaged, one-phonon [23] generalized density-of-states (GDOS), $g^{(n)}(E)$, is related to the measured dynamical structure factor, $S(Q, E)$ from INS by

$$g^{(n)}(E) = A \left\langle \frac{e^{2W(Q)}}{Q^2} \frac{E}{n(E,T) + \frac{1}{2} \pm \frac{1}{2}} S(Q,E) \right\rangle$$

where $A$ is a normalization constant, $2W(Q)$ is the Debye-Waller factor and $n(E,T)$ is the thermal occupation factor (Bose-factor correction) equal to $[\exp(E/k_BT)-1]^{-1}$. The + or - signs correspond to neutron energy loss or gain respectively and the bra-kets indicate an average over all $Q$. It is worth noting that the Bose factor corrected dynamical structure factor $S(Q, E)$ is generally termed $\chi"(Q, E)$, and referred to as dynamical susceptibility.

### III.   COMPUTATIONAL DETAILS

DFT calculations were carried out using the projector-augmented wave potentials within the Vienna *ab initio* simulation package (VASP) [24–29]. The generalized gradient approximation was adopted and both Perdew-Burke-Ernzerhof (PBE) [30,31] and van der Waals'-based (vdW-DF2) [32–35] exchange-correlation functionals were used. In order to account for weak interactions between the chains when using the PBE functional, a Grimme-type van der Waals' (vdW) correction was used; based in the D3 scheme with a Becke-Johnson (BJ) damping function, PBE-D3 (BJ) [36,37]. For all subsequent mentions of this method, the BJ is dropped for simplicity. Structural relaxations were carried out until the residual forces on each atom were below 0.001 eV/Å. An appropriate *k*-point mesh for each structure was generated using the Monkhorst-Pack method [38]. The Gaussian broadening technique was employed, with a smearing width of 0.01 eV.

The direct method was used for phonon calculations performed on a $2 \times 2 \times 2$ supercell [39]. The GDOS was calculated from the partial vibrational density-of-states (PDOS), $g_k(E)$ of the $k^{\text{th}}$ atom, via

$$g^{(n)}(E) = B \sum_k \left( \frac{4\pi b_k^2}{m_k} \right) g_k(E)$$

where $B$ is a normalization constant, $b_k$ is the neutron scattering length and $m_k$ is the mass of the $k^{th}$ atom [22]. The constant $(4\pi b_k^2/ m_k)$ represents the atom's neutron weighting factor which for Cu, Ag, Au, C and N are 0.1264, 0.0463, 0.0393, 0.4622 and 0.8216 barns amu$^{-1}$, respectively.

The thermal behavior and isothermal mode Grüneisen parameters defined as,

$$\gamma_i^T = -\left(\frac{\partial \ln \omega_i}{\partial \ln V}\right)_T$$

where $\omega_i$ is the mode frequency and $V$ is the volume were calculated using the quasi-harmonic approximation (QHA) [40,41]. This approximation takes into account implicit anharmonicity due to changes in volume but neglects the explicit anharmonicity due to temperature. The QHA is valid so long as the explicit anharmonicity does not dominate the total anharmonicity. The phonon density-of-states, mode frequencies and Grüneisen parameters were determined using the Phonopy software [42].

## IV. RESULTS AND DISUCSSION

The synchrotron-based powder x-ray diffraction pattern of $Cu_{1/3}Ag_{1/3}Au_{1/3}CN$, shown in Fig. 2, revealed the compound is single phase and adopts the AuCN structure. The lattice parameters at 100 K obtained from a Pawley refinement of a hexagonal cell are $a = 5.875(1)$ and $c = 15.011(8)$ Å. All known compounds with the AuCN structure exhibit 1D NTE [5]. For example the NTE coefficient, $\alpha_c$, of AuCN and $Cu_{1/2}Au_{1/2}CN$ is -8.9 and -13.8 MK$^{-1}$, respectively [5]. Hence the thermal expansion behavior of $Cu_{1/3}Ag_{1/3}Au_{1/3}CN$ was investigated. The expansion coefficients were extracted from the refined lattice parameters at each temperature over the range 100-500 K, using the gradient of an asymmetric sigmoid fit. The results revealed not only is there no NTE along the chain, but small PTE is observed. Colossal PTE perpendicular to the chains resulting in a large volume expansion is also present. The maximum value of the expansion coefficients $\alpha_a$, $\alpha_c$ and $\alpha_V$ are +85, +18 and +184 MK$^{-1}$, occurring at 238, 333 and 247 K, respectively. The value of $\alpha_a$ is larger than for AuCN (61.6 MK$^{-1}$) but similar to $Cu_{1/2}Au_{1/2}CN$ (83.3 MK$^{-1}$). However the additional PTE along the chain results in a volume expansion larger than both ($\alpha_V$ of AuCN is 115.3 MK$^{-1}$ and $\alpha_V$ of $Cu_{1/2}Au_{1/2}CN$ is 154.6 MK$^{-1}$) [5].

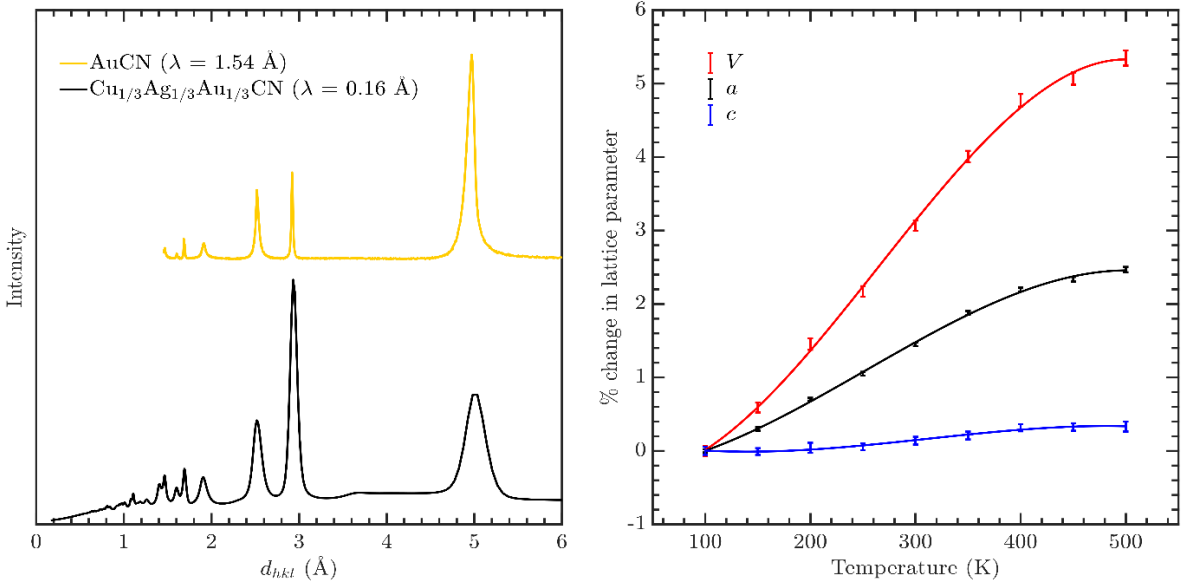

FIG. 2. Left: Synchrotron-based x-ray diffraction (XRD) pattern of $Cu_{1/3}Ag_{1/3}Au_{1/3}CN$ measured at 100 K, compared to that of commercial AuCN measured at room temperature, using a Bruker D8 diffractometer equipped with primary Cu K$_{\alpha 1}$ radiation. $Cu_{1/3}Ag_{1/3}Au_{1/3}CN$ is single phase and clearly adopts the AuCN structure. Right: The percentage change in the lattice parameters with temperature from a Pawley refinement of the synchrotron-based XRD measurements. The points are fitted with an asymmetric sigmoid function.

As in $Cu_{1/2}Au_{1/2}CN$, $Cu_{1/3}Ag_{1/3}Au_{1/3}CN$ lacks long-range ordering between the chains [5]. Hence x-ray diffraction alone is not sufficient to build a structural model for dynamical calculations and further techniques are required. Furthermore, as many transition-metal cyanides have 'head-to-tail' disorder in the relative orientation of the cyanide ligand [2-5], the bonding preference of the carbon was investigated. The direct-polarization magic-angle spinning (DPMAS) $^{13}$C SSNMR spectrum is shown in Fig. 3. The spectrum shows a resonance at 163 ppm with a shoulder at 157 ppm, which can be attributed to $^{13}$C bonded to Au and $^{13}$C bonded to Ag, respectively. The side bands are a consequence of the magic-angle spinning. No doublet or quartet are observed, as the electric-field tensors of Ag and Au are larger compared to their Larmor frequencies. If the $^{13}$C were directly bonded to $^{63/65}$Cu, a multiplet structure would be observed with resonances below 140 ppm [43], which are not present here. Hence C is never bonded directly to Cu, but instead preferentially bonds to Au or Ag. This information additionally provides evidence that there are never two consecutive Cu ions along the chain. Furthermore, qualitative analysis of the NMR peaks, indicates that C bonds more often to Au ions than Ag ions. Combined with the fact that the $c$ lattice parameter from the Pawley refinement is equal to three times that of AuCN, it is reasonable to assume a structural model where the three metals repeat along the chain as Cu-NC-Ag-NC-Au-CN.

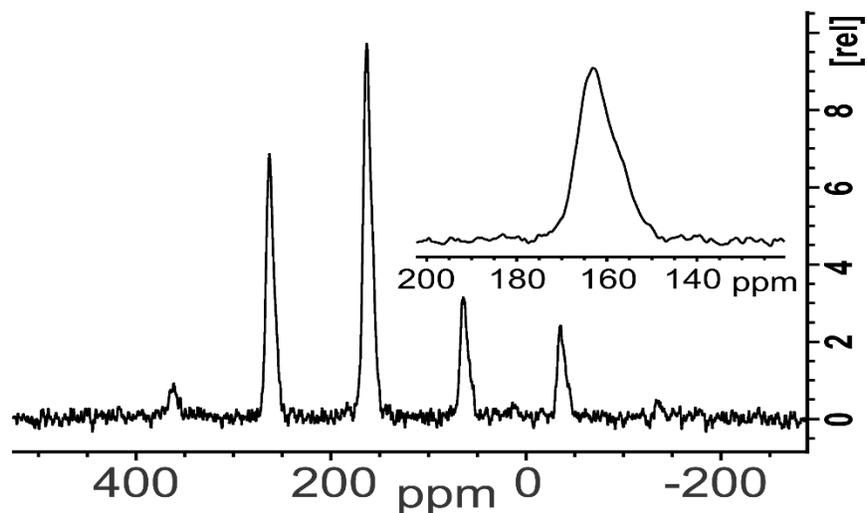

FIG. 3. $^{13}$C DEPTH SSNMR spectrum of $Cu_{1/3}Ag_{1/3}Au_{1/3}CN$. The isotropic peak is shown top right with a main resonance at 163 ppm ($^{13}$C bonded to Au) and a shoulder at 157 ppm ($^{13}$C bonded to Ag). No resonances are observed corresponding to $^{13}$C bonded to Cu.

The PDF, shown in Fig. 4, from the synchrotron-based total x-ray diffraction at 100 K was used to compare with simulated PDFs of various structural models. The first peak at ~2.0 Å represents the metal-to-cyanide bond lengths, whereas the larger peak at ~3.4 Å is the distance between neighboring chains. The small peak at ~5.0 Å is the from the Cu-Ag and Cu-Au distances along the chain, however the Ag-Au distance is longer, resulting in the peak at ~5.2 Å. Three potential models are also shown in Fig. 4 and their fractional coordinates are listed in Table 1. The first, with homometallic layers, clearly does not represent the local order but acts as a good comparison for the other two models. The second model contains heterometallic layers, with each metal having only unlike nearest neighbors. Model 2 has lattice parameters corresponding to those resulting from the Pawley refinement at 100 K. Due to the hexagonal packing of the chains, the $a$ parameter of Model 1 corresponds to $1/\sqrt{3}$ of the $a$ parameter for Model 2. The third model contains homometallic layers of Ag atoms, along with layers containing Cu and Au. Model 3 has orthorhombic rather than hexagonal symmetry with lattice parameters $a$ and $b$, corresponding to the $a$ lattice parameters of Models 1 and 2, respectively.

As expected Model 1, consisting of homometallic layers (Fig. 4), does not reproduce the PDF data well. The intensity of the first peak at 3.4 Å is too high as each Au ion is surrounded by six other Au ions. Models 2 and 3 fit significantly better, although it is difficult to distinguish between the two. It is possible that the real compound contains contributions from both models.

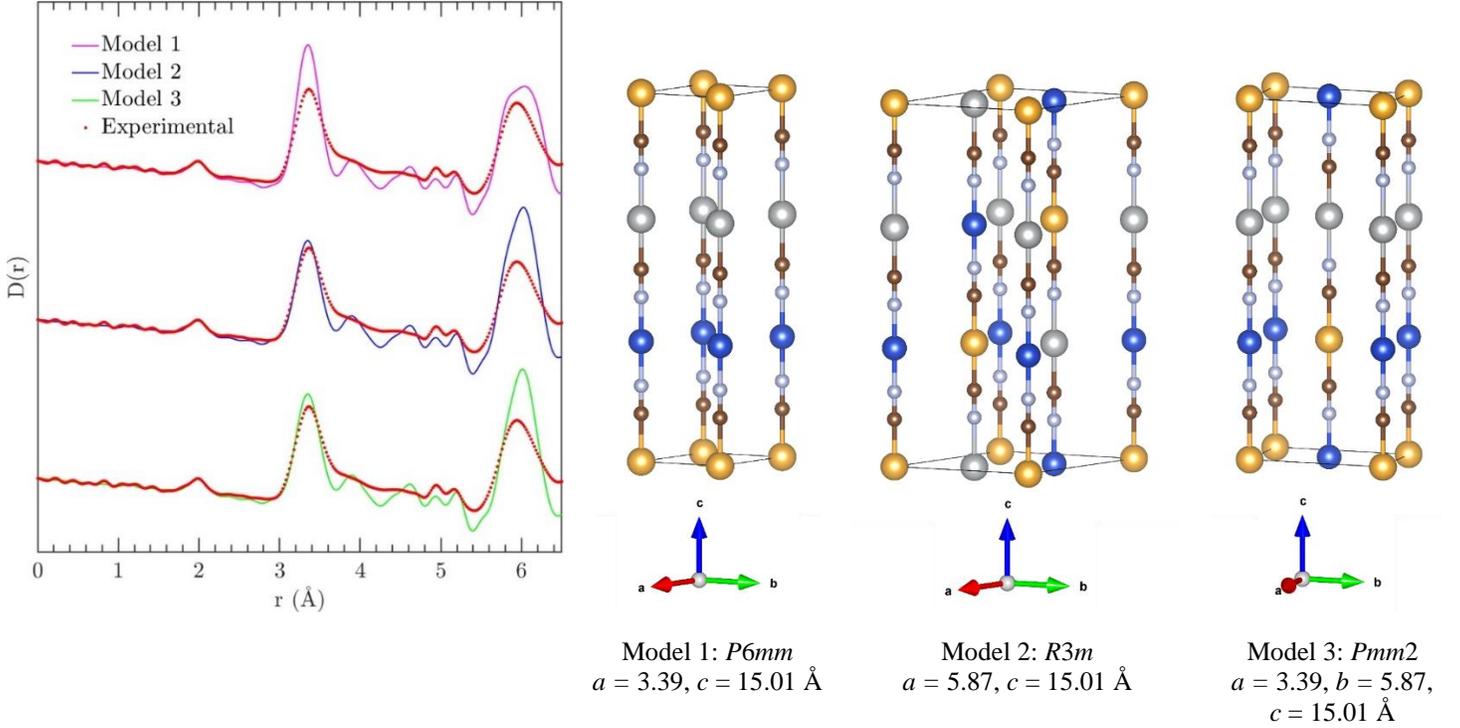

FIG. 4. Left: Comparison of the calculated (solid lines) $D(r)$ of the different structural models with experiment (dotted line). Right: Schematic illustration of the modelled structures, and their associated space groups, used to reproduce the measured PDF data. The structural models consist of: homometallic layers (left), heterometallic layers (center), and a combination of homometallic Ag layers and mixed Cu-Au layers (right). Key: Cu, blue, Ag, silver, Au, yellow, C, brown, N, grey.

TABLE 1. Fractional atomic positions for the three models illustrated in FIG. 4.

| For all models | | | | Additionally for Model 3. | | | |
|---|---|---|---|---|---|---|---|
|  | $x$ | $y$ | $z$ |  | $x$ | $y$ | $z$ |
| Au | 0 | 0 | 0.0000 | Au | ½ | ½ | 0.3260 |
| Ag | 0 | 0 | 0.6580 | Ag | ½ | ½ | 0.6680 |
| Cu | 0 | 0 | 0.3260 | Cu | ½ | ½ | 0.0000 |
| C | 0 | 0 | 0.1301 | C | ½ | ½ | 0.1958 |
| C | 0 | 0 | 0.5218 | C | ½ | ½ | 0.4561 |
| C | 0 | 0 | 0.8699 | C | ½ | ½ | 0.8042 |
| N | 0 | 0 | 0.2058 | N | ½ | ½ | 0.1201 |
| N | 0 | 0 | 0.4461 | N | ½ | ½ | 0.5318 |
| N | 0 | 0 | 0.7942 | N | ½ | ½ | 0.8799 |

Each Model was structurally relaxed using DFT and improved DFT methods (see Section III), to either neglect or include van der Waals' (vdW) interactions, respectively. Figure 5 shows the resulting lattice parameters and total energies, which are also listed in Table 2. For comparison, the lattice parameter perpendicular to the chains of Model 2 is referred to as $b$. For all DFT schemes tested, Model 2 led to the lowest-energy structure. The energy difference between Model 1 (the highest-energy structure and therefore the least stable) and Model 2 was smallest when vdW interactions were not included. This approach also resulted in $a$ and/or $b$ lattice parameters that are significantly larger than the experimental values. As the parameters represent distances between the chains, their inflation is expected when weak attractive forces are not considered. The vdW-DF2 functional, which includes a vdW correction within the exchange correlation potential, resulted in reasonable inter-chain lattice parameters, but overestimated the $c$ lattice parameter. The use of a Grimme-type correction (PBE-D3) however, resulted in lattice parameters reproducing the experimental values at 100 K. Model 3 was also relaxed using the vdW-DF2 scheme with all the CN ligands flipped, such that the C end is preferentially bonding to Cu rather than Au. A significant difference of 1.5 eV per 9 formula units (27 atoms) in the total energy was found, further confirming the established cyanide-bonding pattern. The two other possible versions of Model 3 (one with Cu layers and Ag-Au layers and one with Au layers and Cu-Ag layers) were also investigated, and both were found to be energetically less favorable than the model with Ag-only layers shown in Fig. 4, and hence were not investigated further.

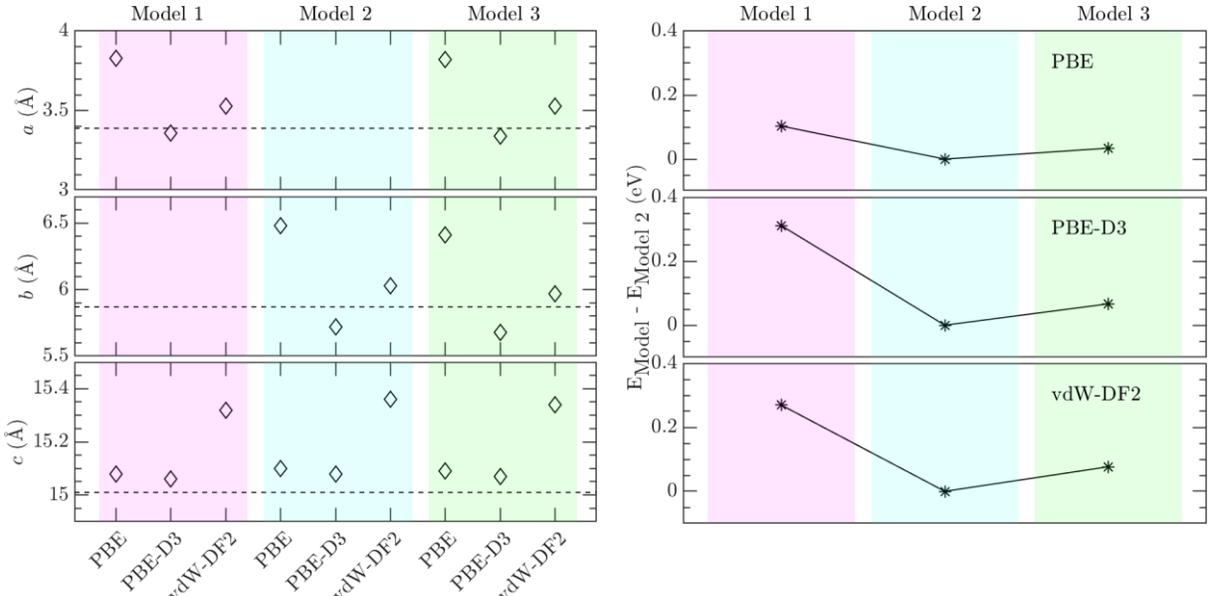

FIG. 5. Left: Relaxed lattice parameters of the different structural models using different DFT schemes, where vdW corrections are either included or neglected. The dotted line represents the experimental lattice parameters extracted from a Pawley refinement of Bragg x-ray data at 100 K. Right: Total energies relative to Model 2 per 9 formula units (27 atoms).

TABLE 2. Total energy (eV) per 9 formula units (27 atoms) of the three models, illustrated in FIG. 4, using different DFT schemes.

|  | Model 1 | Model 2 | Model 3 |
| --- | --- | --- | --- |
| PBE | -177.3767 | -177.4798 | -177.4457 |
| PBE-D3 | -182.1543 | -182.4655 | -182.3988 |
| vdW-DF2 | -121.3434 | -121.6129 | -121.5361 |

The phonon dynamics of $Cu_{1/3}Ag_{1/3}Au_{1/3}CN$ were probed using INS measurements, as described in Section II. The temperature evolution of the phonon spectra is shown in Fig. 6. The first band at ~5 meV, which is sharp and clearly split (band 1), undergoes a softening with temperature as its maximum shifts to lower energy. It is followed by a broader band with a small shoulder (band 2), spanning 10–28 meV, which also softens with temperature. The third band (band 3) undergoes an interesting evolution. At 200 and 300 K, it appears as a broad band with seemingly two contributions. Then at 400 K, the band clearly has two maxima, one at 34 and one at 42 meV. Finally at 500 K, only the second band at 42 meV is distinct. This temperature evolution could be due to a phonon softening, followed by collapse, indicating a change in local symmetry and/or large anharmonicity. The large softening coincides with the positive thermal expansion along the chain, which occurs at its maximum between 300 and 400 K. The fourth band at 54 meV (band 4) also undergoes a small softening, which at such a high energy is more likely to be due to local symmetry changes than pure anharmonicity.

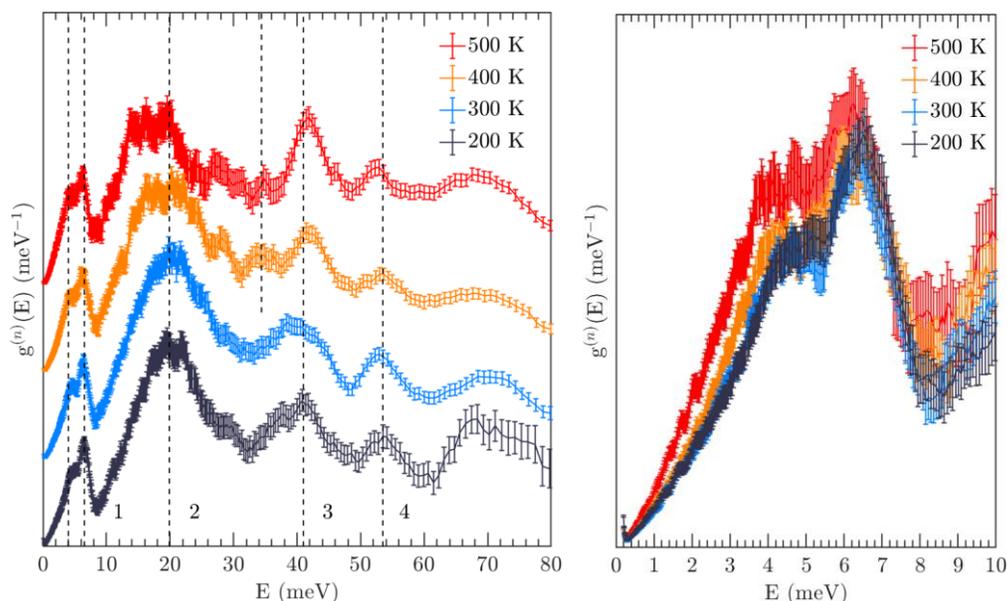

FIG. 6. Left: The temperature evolution of the measured phonon spectra. The vertical lines are a guide for the eyes, and the phonon bands referred to in the text are labelled as 1, 2, 3 and 4. The third band at ~40 meV undergoes an apparent splitting at 400 K with maxima at 34 and 42 meV. The band at 42 meV then sharpens at 500 K, while at 34 meV the band is no longer distinct. Right: A closer view of the first band, which highlights the clear softening.

Phonon calculations were carried out using the three structural models described above, and the resulting phonon spectra for Models 2 and 3 are compared to the experimental measurements (Fig. 7). The results for Model 1 are not shown as they contained imaginary modes indicating that the model is unstable. This is in agreement with the total-energy calculations and the PDF analysis. Very little difference is found between the calculated spectra of Model 2 and Model 3. However, the choice DFT scheme had a large impact on the resulting phonon spectra, especially at low energy. The PBE functional, which neglects vdW interactions and overestimated the interchain lattice parameter, incorrectly predicted both the position and splitting of the first band. The inclusion of vdW interactions, either within the exchange correlation functional (vdW-DF2) or as a Grimme-type correction (PBE-D3), resulted in a significant improvement to the dynamical description of the system, with the PBE-D3 scheme best reproducing low-energy features.

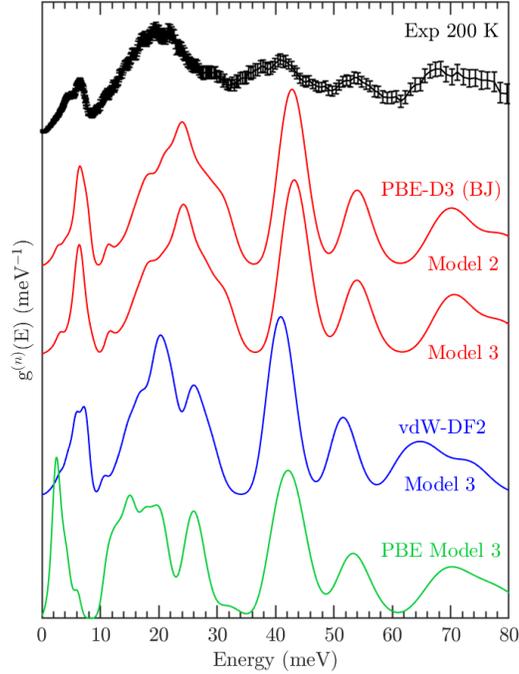

FIG. 7. Comparison of the experimental phonon spectrum at 200 K with the calculated one using different DFT methods and models. Calculations for Model 3 using the PBE-D3, vdW-DF2 and PBE density functional schemes, reveal the inclusion of vdW interactions (PBE-D3 and vdW-DF2) makes a significant difference to the calculated GDOS. Calculations for Models 2 and 3 using the PBE-D3 density functional scheme have little observable difference.

Above 35 meV, the spectrum calculated using the vdW-DF2 functional is red-shifted compared to the one calculated using the PBE functional (without a vdW correction). As modes of this energy are internal vibrations, the shift is due to the longer bond lengths resulting from the higher estimated $c$ parameter of the vdW-DF2 functional. Indeed, the frequency shift is even more apparent on comparison with the experimental infrared and Raman frequencies. The measured optical spectra along with the calculated frequencies of Model 3 ($\Gamma = 17A_1 + 17B_1 + 17B_2$), using both the PBE-D3 and vdW-DF2 scheme are shown in Fig. 8. The calculated frequencies for Model 2 ($\Gamma = 9A_1 + 9E$) are very similar to those for Model 3, as can be seen in the calculated GDOS. Hence, even though the vdW-DF2 functional predicts a large $c$ lattice parameter, it better reproduces the positions of the infrared and Raman peaks.

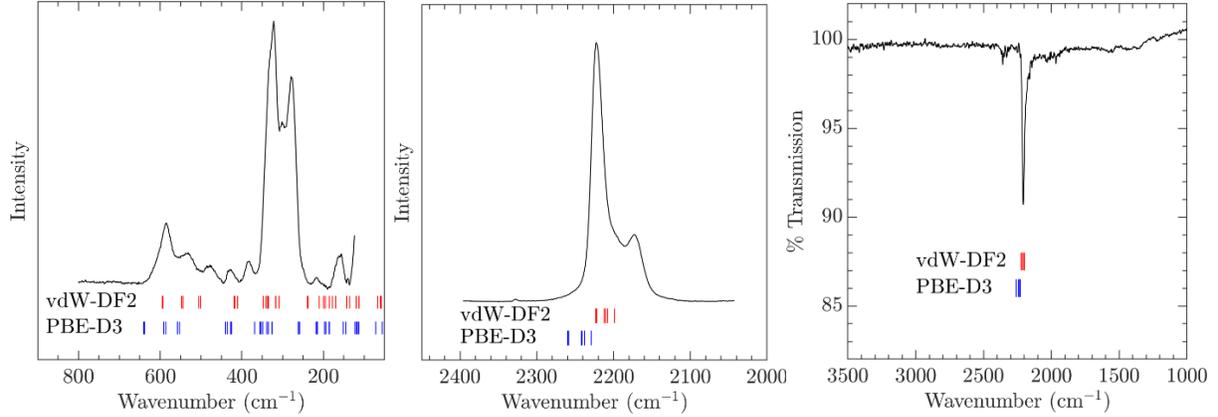

FIG. 8. The measured Raman spectra in the low-energy (left) and CN stretch (center) regions along with the infrared spectrum (right). The calculated frequencies of Model 3 using both the vdW-DF2 and PBE-D3 DFT schemes, accounting for vdW corrections are shown underneath the spectra. The measured and calculated values are listed in Table 3.

TABLE 3. Measured IR and Raman frequencies together with those above 150 cm$^{-1}$ calculated for Model 3. FIG. 8 shows the measured spectra.

| | Experimental | vdW-DF2 | PBE-D3 |
|---|---|---|---|
| $\nu$ (cm$^{-1}$) $150 < \nu < 650$ cm$^{-1}$ | 164, 248, 302, 324, 329, 383, 432, 480, 532, 585 | 170, 170, 178, 186, 196, 200, 210, 238, 238, 239, 308, 309, 317, 317, 335, 337, 341, 347, 410, 411, 417, 419, 501, 505, 544, 548, 593, 595 | 152, 185, 186, 194, 197, 215, 215, 218, 258, 259, 262, 326, 326, 335, 339, 349, 353, 356, 368, 425, 427, 435, 440, 553, 558, 586, 592, 638, 640 |
| $\nu_{C\equiv N}$ (cm$^{-1}$) | 2172, 2206 (IR), 2223 | 2199, 2208, 2210, 2212, 2222, 2223 | 2229, 2237, 2241, 2242, 2258, 2260 |

All the phonon calculations predict the third experimental phonon band at 40 meV to be a single peak, as is observed at 500 K. The modes in this band are rotational motions of the cyanide ligands with the metals at rest. The calculated dispersion curves of Models 2 and 3 (Fig. 9) reveal a small splitting of the phonon modes at this energy when the wave vector lies on the $c^*$=0 plane. The lower-energy modes at ~ 38 meV involve the cyanide ligands oscillating in the same direction, such that the metal ions retain their 180° C/N-M-C/N angle. For the modes at 42 meV however, the angles around M = Cu, Ag become non-linear. A cartoon of this is also shown in Fig. 9. When the wave vector lies on the Brillouin zone face normal to $c^*$, these vibrations have a similar energy as bending will appear in the low-energy modes, and be lost in the higher-energy modes due to the different relative phases along the chain of the cyanide ligand vibrations. Figure 9 also shows the type of vibrations responsible for the fourth experimental phonon band. The motion also consists of localized rotations of the cyanide ligands, but in this case resulting only in a change of the M = Au angle.

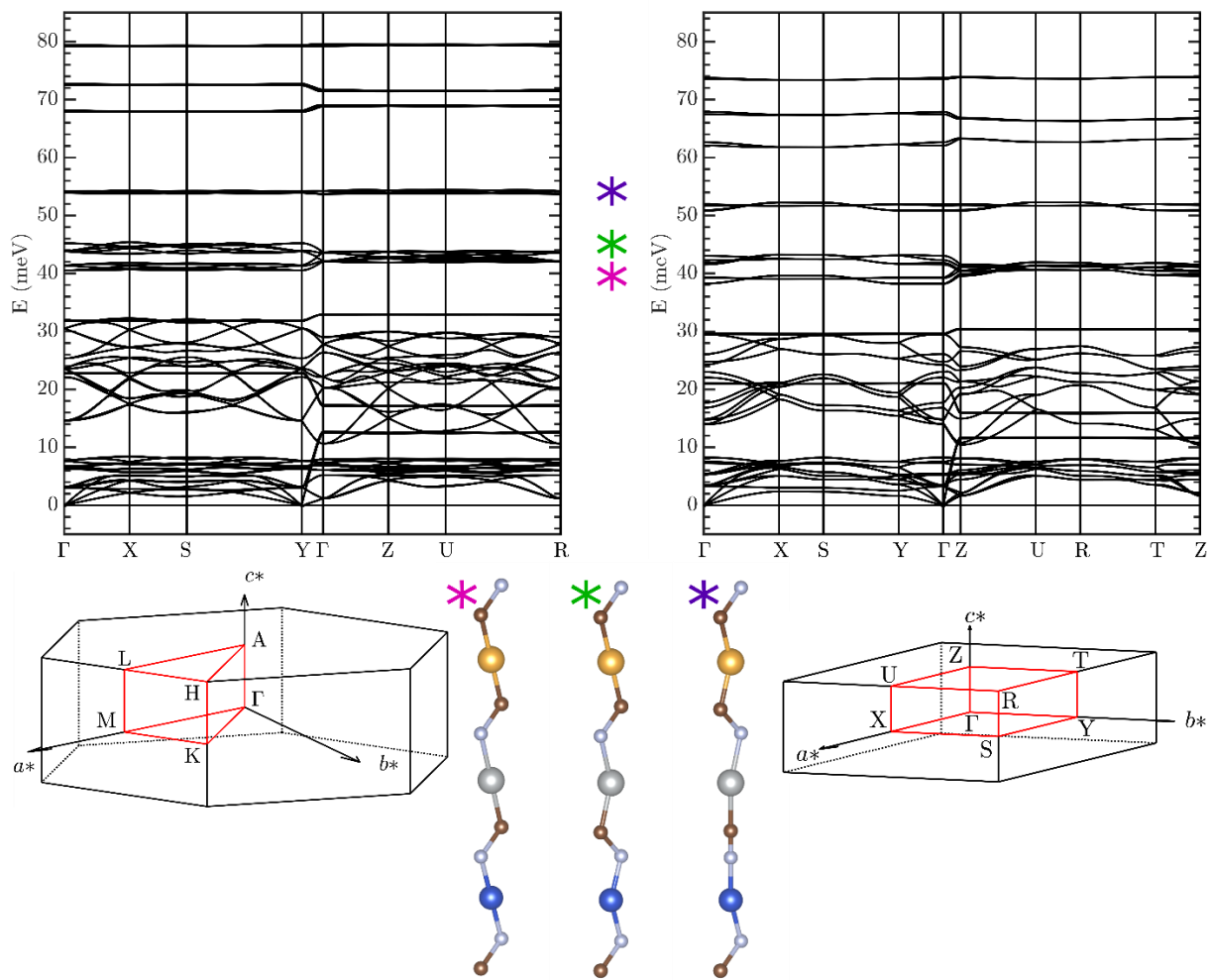

FIG. 9. Phonon dispersion curves of Model 2 calculated using PBE-D3 (left) and of Model 3 calculated using vdW-DF2 (right). Comparison of the two highlights some low-energy differences in the phonon band structure, along with the noticeable frequency shift between the vdW-DF2 and PBE-D3 calculations above 35 meV. The CN stretching modes at ~270 meV are not shown; however, their Γ-point frequencies can be seen in Fig. 8. Underneath the dispersion curves are the first Brillouin zone of each model, with labelled high-symmetry points for Model 2 with *R*3*m* symmetry (left) and Model 3 with *Pmm*2 symmetry (right). The colored asterisks correspond to schematic illustrations of the different localized cyanide-rotation modes for a single chain occurring at ~38, 42 and 52 meV, respectively. In the pink mode all the metal atoms retain their linear connections to C/N, whereas in the green mode, the Cu and Ag configuration becomes bent. As shown, these two vibrations clearly have different energies. However, once the wave vector possesses a *z*-component (along the chain), the motions of the cyanide ligands along the chain will have a different relative phase and hence, for a single chain, these two types of motion will resemble one another more and have similar energies. In the purple mode, only the cyanide ligands bonded to Au are rotating resulting in a bend in the Au connectivity and hence this mode is higher in energy. Key: Cu, blue, Ag, silver, Au, yellow, C, brown, N, grey.

The calculated thermal ellipsoids at 300 K are shown in Fig. 10. For Model 2, the Cu and Ag ions have oblate spheroids, which reveal they have greater displacement perpendicular to rather than along the chain. This is consistent with chain-undulation modes, which induce PTE perpendicular to the chain but also NTE along the chain. On the other hand, C, N and Au have spherical ellipsoids, indicating an equal mean-square displacement in all directions. The greater displacement perpendicular to the chain of Cu and Ag compared to Au provides an explanation as to why the addition of Cu or Ag to AuCN (to form $Cu_{1/2}Au_{1/2}CN$ or $Ag_{1/2}Au_{1/2}CN$) results in greater PTE and NTE due to chain undulation modes [5]. For Model 3, which contains a homometallic layer of Ag, the Ag atoms have a spherical ellipsoid similar to that of Au, implying less undulation. On the other hand, the Cu atoms have an even more exaggerated oblate spheroid than in Model 2. Due to the orthorhombic symmetry of Model 3, the calculated thermal ellipsoid for Cu is actually longer along $b$ than $a$. Hence, the orthorhombic symmetry of Model 3 strongly affects the calculated dynamics.

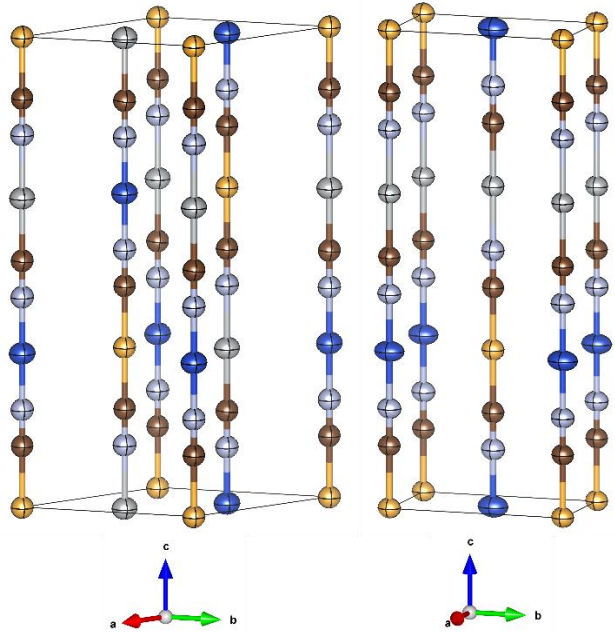

FIG. 10. Calculated thermal ellipsoids at 300 K for Model 2 (left) and Model 3 (right). Key: Cu, blue, Ag, silver, Au, yellow, C, brown, N, grey. The ellipsoid represents 75% probability of containing the atom. For Model 2, both Cu and Ag have slightly oblate spheroids, indicating greater displacement perpendicular to the chain rather than along it. In Model 3 however, only Cu has a highly exaggerated oblate spheroid that also has greater mean-square displacement along $b$ than $a$. Key: Cu, blue, Ag, silver, Au, yellow, C, brown, N, grey.

The volume expansion coefficients, calculated using the quasi-harmonic approximation (QHA) of Models 2 and 3 with the PBE-D3 and vdW-DF2 schemes are show in Fig. 11. Model 2 is found to better reproduce the measured data, following the correct trend and reaching the same maximum value when calculated using the PBE-D3 DFT scheme, but with a temperature shift. The comparison between the measured linear thermal expansion coefficients and those of Model 2 calculated using PBE-D3 are also shown in Fig. 11. Again, the temperature shift is apparent. Importantly, in all cases no NTE is predicted along the chain. Instead, PTE is predicted in agreement with observations, albeit with a smaller magnitude.

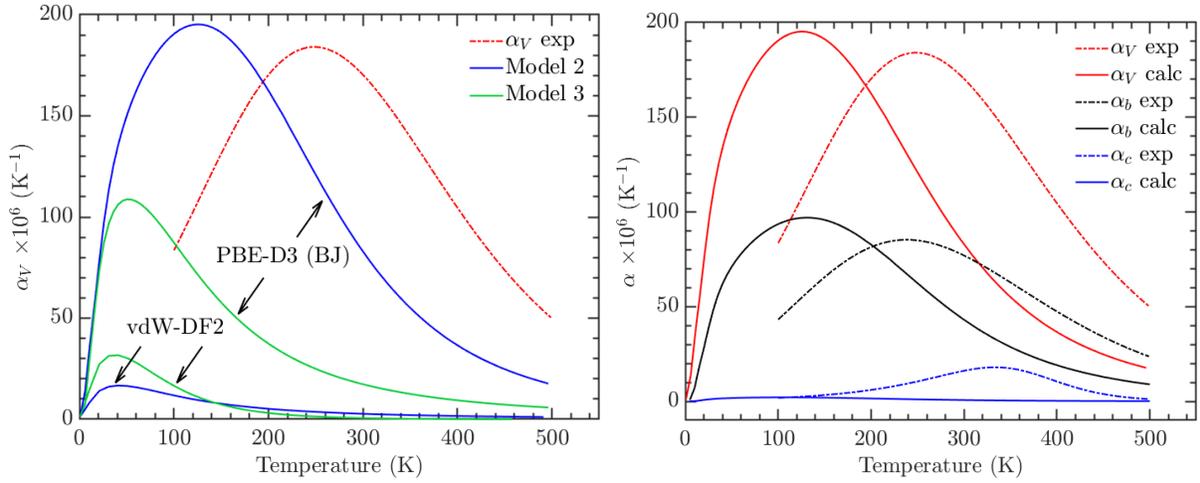

FIG. 11. Left: The experimental and calculated volume expansion behavior. The experimental PTE behavior (red dotted line) is extracted from the fit of the lattice parameters (Fig. 2). The volume expansion of Model 2 calculated using the PBE-D3 scheme best reproduced the observation, but with a temperature shift. Using the PBE-D3 vdW scheme leads to a better agreement with the experimental behavior than vdW-DF2. Right: The calculated thermal expansion coefficients for Model 2 using the PBE-D3 method (solid lines) compared to the experiment coefficients (dotted lines).

The mode Grüneisen parameters of Model 2, calculated using the PBE-D3 vdW scheme, are shown in Fig. 12. These reveal that modes in the sharp first band and broad second band are responsible for the observed volume PTE. Specifically the optic modes between 5 and 9 meV make the largest contribution to the positive volume expansion. Analysis of the mode eigenvectors reveals that these are chain undulation modes consisting of transverse motions of the metal ions. These modes, however, are known to exist in all group 11 cyanides and are expected to lead to NTE along the chain.

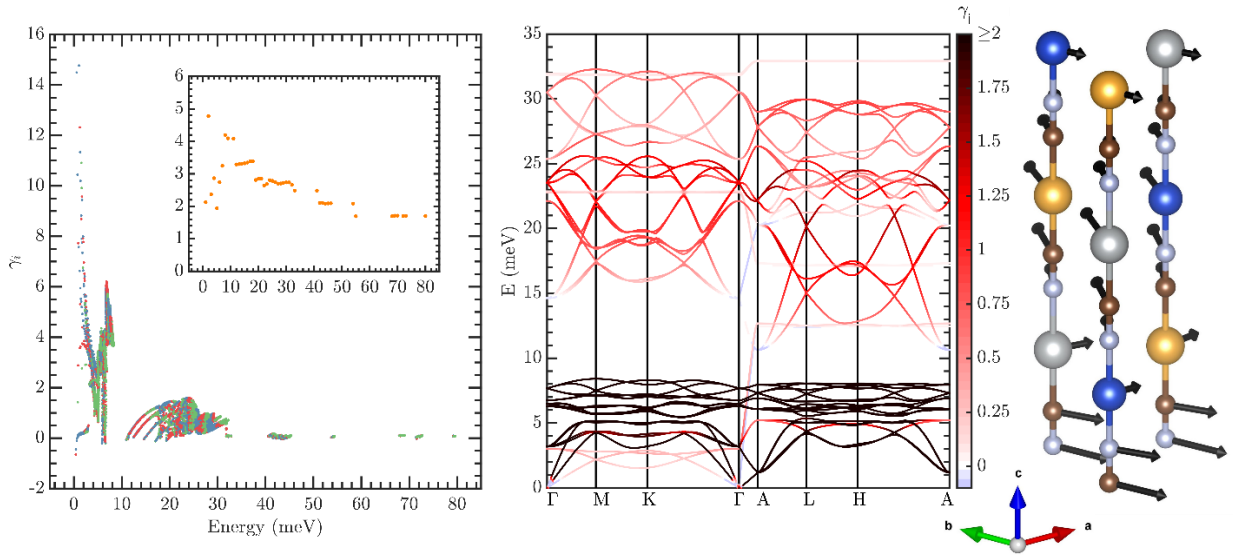

FIG. 12. Left: The calculated mode Grüneisen parameters of Model 2. Phonons are colored differently for contrast. The inset shows the average Grüneisen parameter for modes of a certain energy E. Center: The calculated dispersion curves colored according to the Grüneisen parameters. Right: The real displacement vectors of the lowest-energy acoustic mode at the A high-symmetry point, which has the largest Grüneisen parameter. Key: Cu, blue, Ag, silver, Au, yellow, C, brown, N, grey.

In order to identify the modes driving the calculated PTE along the chain, further analysis of the mode eigenvectors was carried out. All modes with motion in the $c$ direction are shown in Fig. 13. Given that the calculations predict a very small expansion of $c$, the associated Grüneisen parameters are also expected to be small. One acoustic mode and two optic modes with energies below 3 meV were found to have Grüneisen parameters of ~0.5. These modes are chain-slipping motions.

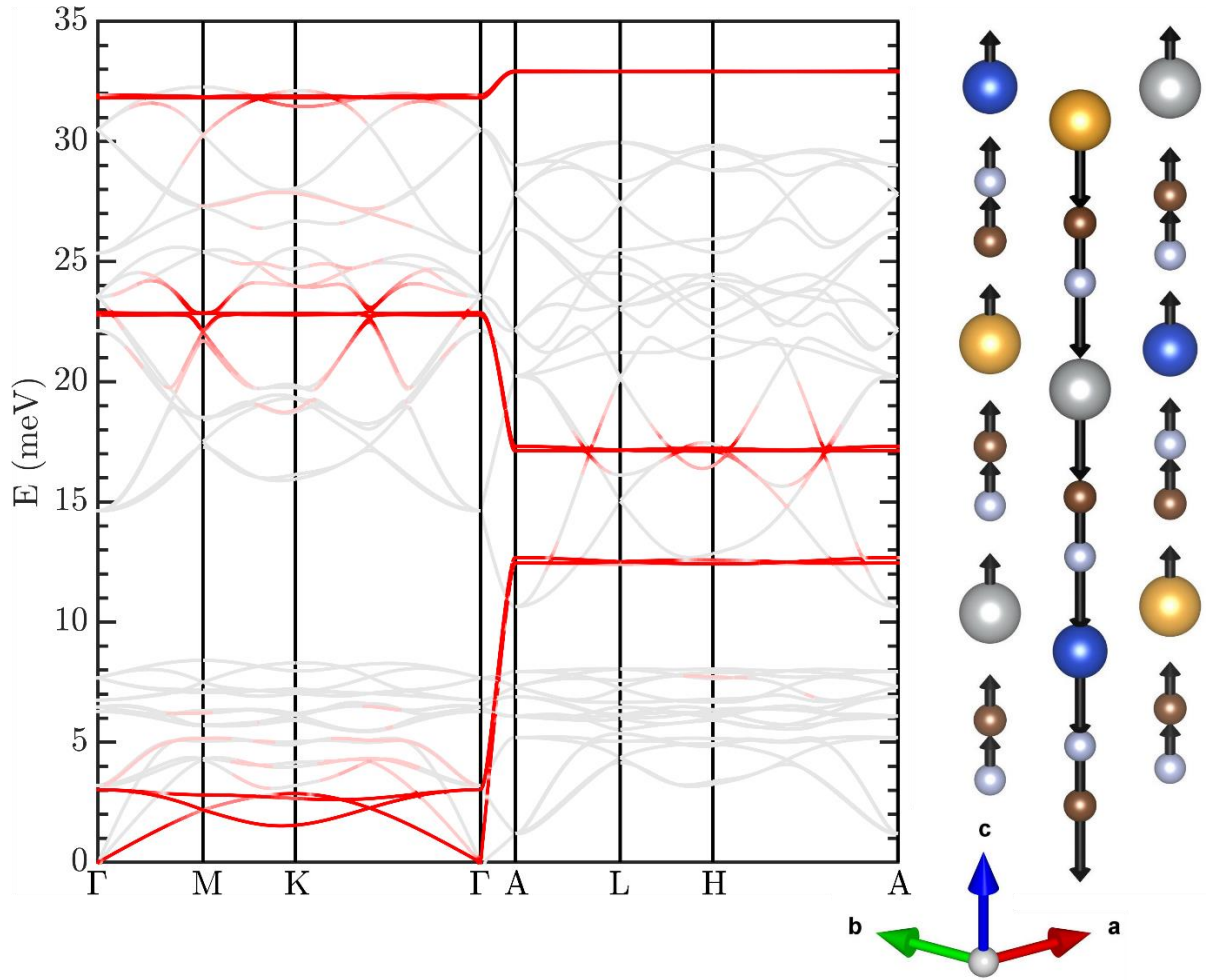

FIG. 13. Left: Phonon modes with a z-component in the displacement vectors. The three lowest-energy modes (1 acoustic and 2 optic) visible on the Γ-M-K-Γ path are chain-slipping modes with a Grüneisen parameter of ~0.5. These are likely the modes responsible for the PTE along the chain. Right: The real displacement vectors of the lowest-energy optic mode at the Γ point. This mode represents a perfect chain-slipping motion. Key: Cu, blue, Ag, silver, Au, yellow, C, brown, N, grey.

Similar chain-slipping motions must exist in AuCN [2], so their positive thermal expansion effect in $Cu_{1/3}Ag_{1/3}Au_{1/3}CN$ is presently intriguing. In MCN and $M_{1/2}M'_{1/2}CN$ (M, M' = Cu, Ag, Au) compounds, there is only one metal-to-metal distance along the chain, provided the metals alternate along the chain. In other $M_xM'_{1-x}CN$ compounds, there are two such distances. However, in $Cu_{1/3}Ag_{1/3}Au_{1/3}CN$ there are three separate metal-to-metal distances along the chain. Hence, as Model 1 clearly does not represent the local order, a maximum of only one in every three metal planes can truly be coplanar. Indeed, structural relaxation of both Models 2 and 3 resulted in none of the metal layers being fully co-planar. In Model 2, a maximal difference of 0.15 Å in the $z$-position of neighboring metal atoms was found and in Model 3, the difference was 0.19 Å for the Ag layer and 0.07 Å for the Cu-Au layer. Hence the metallophilic interactions between neighboring metals, if imagined as weak bonds, are not perfectly perpendicular to the chains themselves. In this case chain-slipping modes result in a small *bond-rotation effect* [44], which then leads to positive thermal expansion along the chain.

The contribution to the mean-square displacement (MSD) with energy of each ion, projected along $a$ and $c$ is shown in Fig. 14. Here it is clear that the modes between 5 and 9 meV, which contribute the most to the PTE perpendicular to rather than along the chains, consist of transverse motions of the metal ions. The majority of contributions to the MSD along $c$ occur below 5 meV and are almost the same for all the ions, consistent with chain-slipping modes. The change in MSD along $a$ and $c$ with temperature, also shown in Fig. 14, highlights the greater displacement of the Cu and Ag ions perpendicular rather than along the chain.

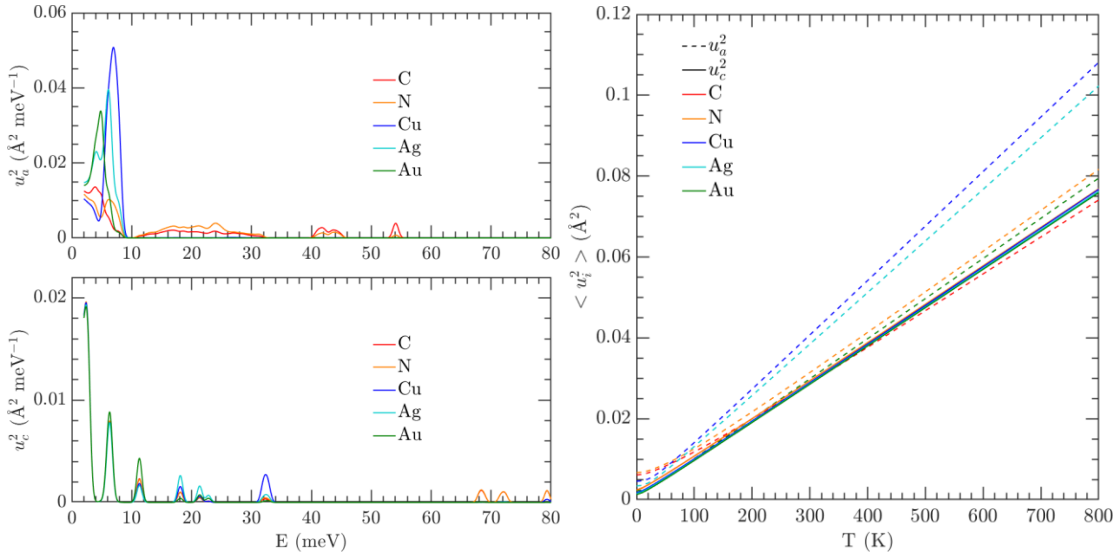

FIG. 14. Left: The contribution at 300 K to the mean-square displacement related to phonon of energy, E, along the $a$ (top) and $c$ (bottom) crystallographic axes for each atom. Here it is clear that the modes between 5 and 9 meV, which cause the PTE perpendicular to the chains, stem from transverse motions of the metals. However the lowest energy modes are dominated by motions of the atoms along the chain. Note the scale for the two directions is different, highlighting the greater displacement perpendicular to the chain of the metals, especially for Cu. Right: The thermal displacement along the two crystallographic axes for each atom with temperature. The displacement along $c$ is almost identical for all atoms. However the transverse motion perpendicular to the chains of the Cu and Ag ions is significantly larger than the other atoms.

The PTE along the chain predicted using the QHA is not of same magnitude as that observed experimentally. As the QHA neglects explicit anharmonic effects, we extracted the total anharmonicity of $Cu_{1/3}Ag_{1/3}Au_{1/3}CN$ from the cumulative GDOS measurements at 200 and 300 K. The total anharmoncity together with the implicit anharmonicity calculated via the QHA is shown in Fig. 15. Above 1 meV, the two are well matched although a discrepancy is observed below this energy. It is therefore likely that the very low-energy acoustic modes below 1 meV have large explicit anharmonicity.

There is another plausible explanation for the lack of NTE in $Cu_{1/3}Ag_{1/3}Au_{1/3}CN$; the chains are not completely straight. In this case, the *tension effect* from transverse motions would be mitigated by small structural rearrangements, as discussed in Barron *et al.* [46]. Additionally, if as temperature increases the chains straighten, PTE along the chain would occur. The straightening would also explain the unusual evolution of the third phonon band. It also justifies why the phonon spectra calculated with a straight-chain model better resemble the measured data at 500 K.

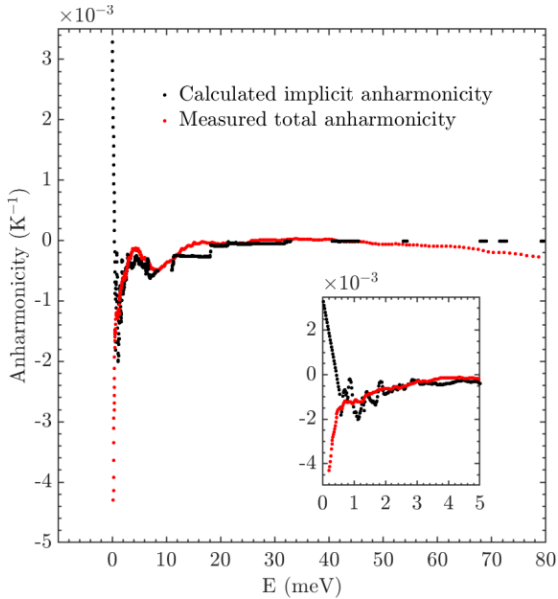

FIG. 15. The measured total and calculated implicit anharmonicity. The difference represents the explicit anharmonicity. Clearly, large anharmonicity is present at very low energy, which is not predicted by the quasi-harmonic approximation. Hence, it is likely that the low-energy chain-slipping modes have large explicit anharmonicity.

## V. CONCLUSIONS

The quasi-1D mixed-metal cyanide, $Cu_{1/3}Ag_{1/3}Au_{1/3}CN$, was synthesized as a single phase adopting the AuCN structure. The compound exhibits large PTE perpendicular to the chains and unlike other related group 11 cyanides, small PTE along the chains as well. Synchrotron-based PDF analysis and SSNMR was used to help model the structure of the compound, and three models were built accordingly. DFT-based structural relaxations ruled out Model 1, and pointed successfully towards the two other models, Model 2 and Model 3, with Model 2 being slightly energetically more stable than Model 3. Due to a lack of long-range metallic order, the real structure is likely to include combinations of both Models 2 and 3.

Variable-temperature INS measurements revealed large anharmonicity in phonon modes below 10 meV, and highlighted the specific evolution of a phonon band at ~ 40 meV, pointing towards possible changes in the local symmetry. Comparison of the implicit anharmonicity calculated using the QHA and total anharmonicity extracted from the INS measurements showed that very low-energy acoustic modes exhibit a pronounced explicit anharmonicity. Phonon calculations, which included vdW corrections using different DFT schemes, reproduced the measured INS spectra well. This highlights the importance of metallophilic interactions between the chains allowing a better description of the observed dynamical behavior.

Mode Grüneisen parameters and thermal expansion coefficients were calculated within the QHA framework. The results for Model 2 reproduced the large PTE perpendicular to the chains, giving both the same maximum value and trend as the experimental data, but with an apparent temperature shift. Analysis of mode eigenvectors revealed chain undulation modes between 5 and 10 meV are responsible for the large PTE perpendicular to the chain, as previously seen in other group 11 cyanides. These modes lead to the *tension effect*, however the calculations also predicted small PTE along the chains. Specific modes below 3 meV consisting of chain-slipping motions were found to have isothermal Grüneisen parameters of ~ 0.5. These modes predict a PTE effect along the chain due to the lack of true metal co-planarity in Model 2, which affects the nature of the metallophilic interactions.

As the PTE thermal expansion along the chain occurs with a greater magnitude than that predicted by DFT, another expansion mechanism is likely occurring. It is possible that small bends in the chains exist to make the metals more co-planar than they would otherwise be. In this case, to compensate for the different metal-to-cyanide bond lengths, the cyanide ligands are not truly parallel to *c*. At low temperatures, the non-linear bonding geometries in the chain mitigate the *tension effect* caused by transverse motions of the metals that induce the PTE perpendicular to the chain. As temperature increases and the chains move further apart, the metallophilic interactions weaken, straightening the chains. This leads to positive thermal expansion and a change in local symmetry as suggested by the evolution of the phonon spectra around 40 meV.

The origin of the positive thermal expansion along the chain in $Cu_{1/3}Ag_{1/3}Au_{1/3}CN$ is clearly routed in the metallophilic interactions that stabilize the AuCN structure. These interactions are disturbed in $Cu_{1/3}Ag_{1/3}Au_{1/3}CN$ compared to other 1D cyanides, due to the different metal-to-cyanide bond lengths along the chain leading to the observed unexpected thermal behavior.

## ACKNOWLEDGEMENTS

The ILL is thanked for providing beam time on the IN6 spectrometer for the inelastic neutron scattering measurements. The x-ray measurements were carried out with the support of the Diamond Light Source, instrument I15-1. The use of the Chemical Analysis Facility (CAF) at Reading and the NMR Facility at Oxford is acknowledged. S.d'A. thanks the ILL and the University of Reading for a PhD studentship.